\documentstyle[prb,aps,multicol,epsfig]{revtex}

\title{The Amorphous-Crystal
       Interface in Silicon: a Tight-Binding Simulation}
\author{N. Bernstein$^{a}$, M.~J. Aziz$^{a}$ and E. Kaxiras$^{a,b}$ }
\address{
$^{a}$Division of Engineering and Applied Sciences, Harvard University,
Cambridge, MA 02138 \\
$^{b}$Physics Department, Harvard University, Cambridge, MA 02138
}

\begin{document}

\maketitle

\begin{abstract}

The structural features of the interface between the cystalline and
amorphous phases of Si solid are studied in simulations based on a
combination of empirical interatomic potentials and a nonorthogonal
tight-binding model.  The tight-binding Hamiltonian was created and
tested for the types of structures and distortions anticipated to occur
at this interface.  The simulations indicate the presence of a number
of interesting features near the interface.  The features that may lead
to crystallization upon heating include $\langle 110 \rangle$ chains
with some defects, most prominently dimers similar to those on the
Si(001) $2 \times 1$ reconstructed free surface.  Within the amorphous
region order is lost over very short distances.  By examining six
different samples with two interfaces each, we find the energy of the
amorphous-crystal interface to be 0.49 $\pm$ 0.05 J/m$^2$.

\end{abstract}

\pacs{}

\begin{multicols}{2}

\section{Introduction}

The crystalline and amorphous phases of silicon are considered
prototypical examples of a tetrahedrally coordinated network in ordered
and disordered forms.  Each phase has been intensively studied
experimentally and theoretically, and both are used in a very broad
spectrum of electronic applications.  Currently all Si integrated
circuits employ several ion implantation steps in their fabrication.
Regions that receive a sufficiently high implantation dose become
amorphous within approximately 100 nm of the free surface; the crystal
structure is restored by an interface-mediated crystallization process
called solid-phase epitaxial growth (SPEG).  While much is known about
the structure of the crystal and amorphous phases individually,
considerably less direct information is available about the structure
of the interface between them.  In light of the importance of SPEG, and
of the intrinsic interest of interfacial phenomena, a detailed
atomistic study of the amorphous-crystal interface in Si is
appropriate.  The inaccessibility of the interface atomic structure by
experimental probes leaves as the only alternative realistic
simulations of this system.  In this paper we discuss such simulations
and the insight they provide into the atomic structure and dynamics at
the amorphous-crystal interface in Si.

There are two major obstacles in simulating this interface:  first, a
relatively large number of atoms must be included in the simulation to
ensure that the character of the two phases is represented accurately;
second, extensive exploration of configuration space is required to
ensure that the system is not locked in some small (and potentially not
representative) region of the accessible configuration space.  Ideally
one would like to simulate this system by means of unbiased,
parameter-free quantum mechanical calculations (such as density
funtional theory in the local density approximation - DFT/LDA), but
both the size of systems that can be handled and the extent of
configuration space that can be explored through such calculations are
severely restricted.  Past efforts have either used hand-built
models\cite{Spaepen78,SaitoOhdomari8184}, computer relaxed geometrical
models\cite{Wooten89}, or Molecular Dynamics (MD) simulations based on
empirical interatomic potentials\cite{Erkoc87,Gilmer94}.  

Here we combine the use of the Stillinger-Weber empirical interatomic
potential and a recently developed semi-empirical quantum mechanical
technique, based on a nonorthogonal tight-binding (TB) Hamiltonian
which was parametrized to represent accurately a wide range of bulk and
surface structures of Si~\cite{BernsteinKaxiras}.  The use of the
empirical potential was motivated by the fact that it affords fast but
less accurate calculations for parts of the simulation where
maintaining high accuracy is not important; specifically it used to
bring the system from a high-temperature, liquid-crystal interface far
from equilibrium, to a low temperature amorphous-crystal interface near
equilibrium.  Once the system is close to equilibrium, we switch to the
TB model which can handle reasonably large systems and is sufficiently
fast to allow exploration of configuration space, while maintaining the
basic quantum mechanical treatment of electronic degrees of freedom.
As such, it is superior to empirical interactions which are
considerably more restricted in their ability to describe large
structural distortions and the breaking and formation of covalent
bonds.  The results of the tight-binding studies can also be used as
starting points for more elaborate parameter-free quantum mechanical
calculations, although we anticipate that the essential features will
remain unchanged.

\section{Methodology} 

We use constant temperature, constant stress MD to prepare the
amorphous-crystal interface samples starting with a liquid-crystal
interface as described below.  Because of the large time scale
necessary to create reasonably equilibrated amorphous samples, the use
of the tight-binding Hamiltonian to compute the interatomic forces
while the system is very far from its equilibrium state is impractical,
and indeed not beneficial.  Instead we use the Stillinger-Weber
interatomic potential\cite{StillingerWeber85} to bring the system
reasonably close to equilibrium and only then turn on the tight-binding
Hamiltonian.  The equations of motion are integrated with a Gear
predictor-corrector algorithm\cite{AllenTildesley} with a time step of
1~fs.  The temperature is kept constant using a velocity rescaling
algorithm where the atomic velocities are uniformly scaled to give the
desired temperature once every 500 time steps.  Zero stress is
maintained with an extended system Parrinello-Rahman
approach\cite{ParrinelloRahman}.

\begin{figure}[p]
    \begin{center}
    \epsfig{file=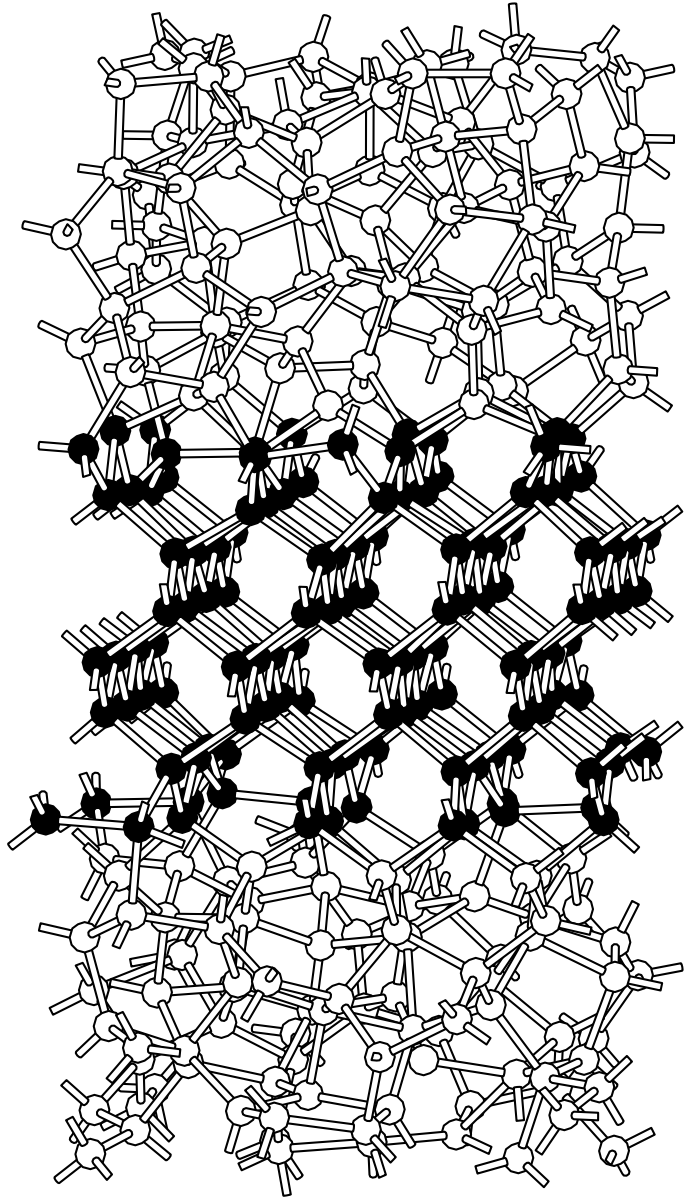,height=6.5in}
    \end{center}
    \begin{center}
    \begin{minipage}[t]{5in}
    \caption{View of a sample along a $(110)$ axis of the crystal.
    Atoms which were kept cool throughout the simulation (corresponding
    to the crystalline region) are in black, atoms in the region that
    was heated then cooled (corresponding to the amorphous region) are
    in white.  Bonds are drawn between atoms closer than a distance of
    2.7~\AA.  Periodic boundary conditions apply in all three
    directions.}
    \end{minipage}
    \end{center}

    \label{fig:sample}
\end{figure}

The simulation cell includes 320 atoms in a
$[220]\times[2\bar{2}0]\times[005]$ box with periodic boundary
conditions in all three directions.  The interface lies in the $(001)$
plane, with the $[220]$ and $[2\bar{2}0]$ vectors forming its sides (in
the following the $[001]$ direction is referred to as the $z$-axis).
To create the interface, we maintain a portion of the simulation cell
in the crystalline phase by keeping it below the melting point, while
melting and then quenching the remainder of the cell.  The crystalline
region includes 128 atoms (8 monolayers) which are kept at 100~K, and
the remaining 192 atoms (12 monolayers) form the amorphous region.  The
amorphous region is produced by cooling a molten region from 5000~K to
1000~K, using the method of Luedtke and Landman\cite{LuedtkeLandman}.
A total of six samples, cooled to 100~K and equilibrated with the
Stillinger-Weber potential, were relaxed using a conjugate-gradient
algorithm with the tight-binding Hamiltonian to calculate forces and
stresses.  The relaxed samples were used in the analysis of interface
features.  A typical sample is shown in Fig. \ref{fig:sample}.

\section{Analysis}

\subsection{Structural analysis}

\begin{figure}[p]
    \begin{center}
    \epsfig{file=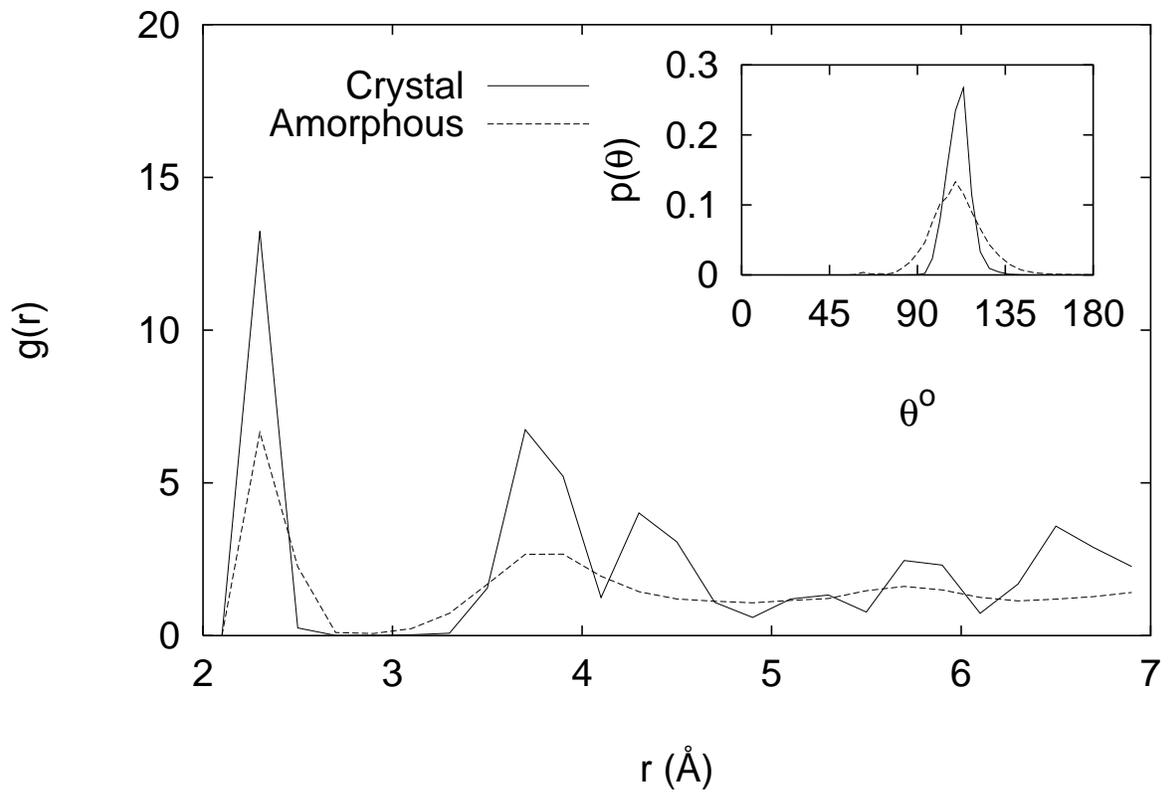,angle=-90,width=6in}
    \end{center}
    \begin{center}
    \begin{minipage}[5]{5in}
    \caption{Measures of order in the bulk of the crystalline and
    amorphous regions (as defined in the text): (a) Pair correlation
    functions $g(r)$ and (b) bond angle distribution functions
    $p(\theta)$. }
    \end{minipage}
    \end{center}

    \label{fig:pc_bad}
\end{figure}

Standard measures for characterizing the structure of the bulk phases
are the radial pair correlation $g(r)$ and bond-angle distribution
$p(\theta)$ functions.  These are shown in Fig.~\ref{fig:pc_bad}
averaged over all six samples.  For the amorphous regions (those that
were thermally cycled), the functions were computed from samples where
the atoms in the crystal regions (those that were kept cold) were
removed, but using the original periodic boundary conditions.  For the
crystal regions the atoms in the amorphous regions were removed.
Because of the missing neighbors at the edges of each region the
normalization for the curves is non-standard, although the overall
shape is not affected.  The pair correlation functions exhibit the
expected features:  averages over atoms in the ``crystalline'' regions
show distinct order at all ranges allowed by the size of the simulation
cell; averages over atoms in the amorphous regions show distinct first
and second neighbor peaks, but no order at longer range.  In
particular, they do not have a third neighbor peak, a feature also seen
in DFT/LDA simulations~\cite{Stich} and in experiment~\cite{Kugler89}.
In the following, the position of the minimum between the first two
peaks of the pair correlation function ($r = 2.7$~\AA) is used as the
criterion for defining the neighbors of an atom in the amorphous
regions.  The mean bond angles are $108.9 \pm 6^\circ$ and $108.4 \pm
14^\circ$ in the crystalline and amorphous regions, close to the ideal
tetrahedral angle of $109.5^\circ$.

Coordination statistics and ring statistics based on the same
nearest-neighbor criterion are listed in Table~\ref{table:bulk_stats}.
The coordination of the atoms in the crystalline region is nearly
perfect; in the amorphous region there is a significant number of
defects, with over-coordinated atoms predominating.  There is also a
significant number of minimal rings (computed using shortest path
analysis~\cite{King67}) with size other than six, including a few
eight-membered rings.  In agreement with the results of DFT/LDA MD
simulations by Stich {\it et al.}~\cite{Stich}, we also observed more
5-membered than 7-membered rings.  The total ring statistics indicate
more even-membered rings than the random bond switching model of Wooten
{\it et al.}~\cite{WWW85,WW87}, and fewer odd-membered rings.

\begin{figure}[p]
    \begin{center}
    \epsfig{file=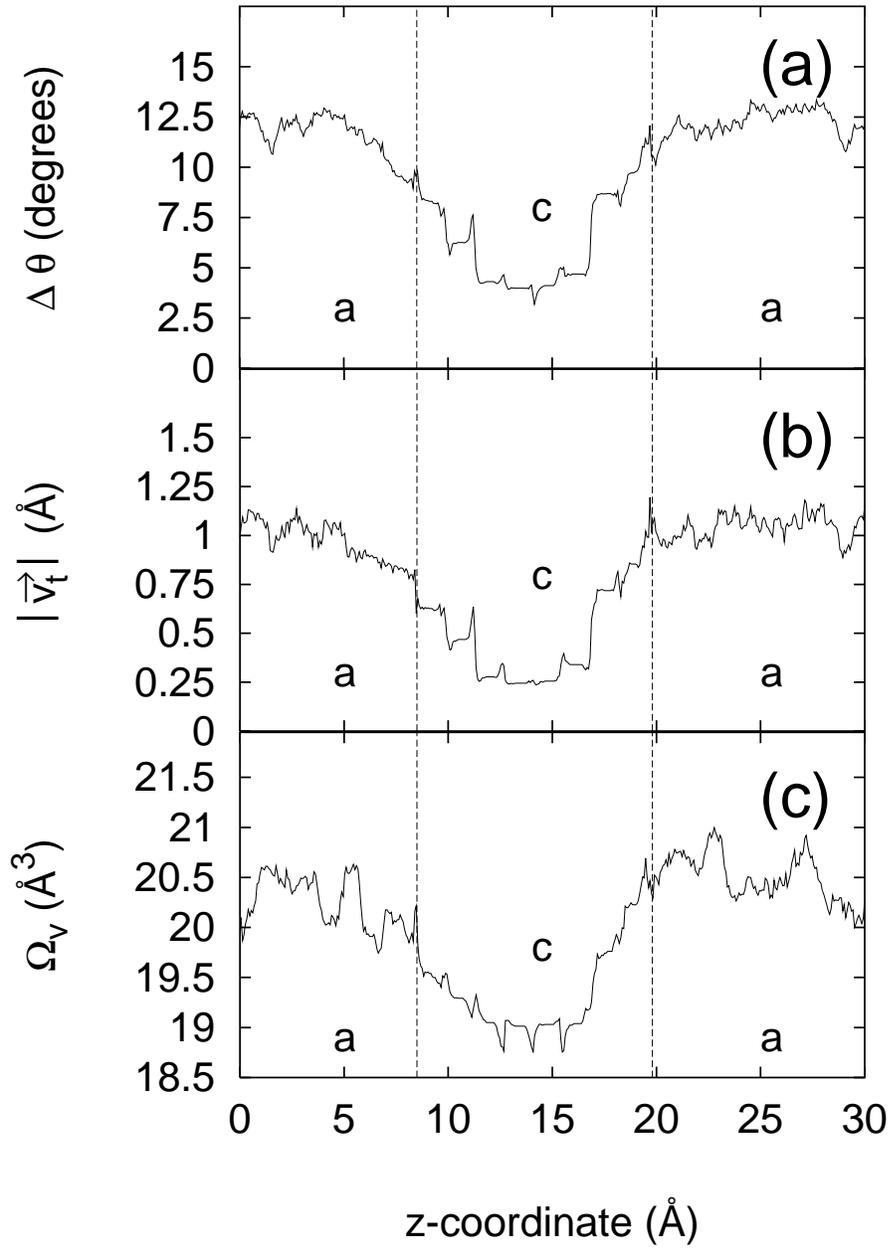,height=6.5in}
    \end{center}
    \begin{center}
    \begin{minipage}[t]{5in}
    \caption{Local measures of order through the samples containing
    a crystal-amorphous interface, averaged over six samples.  The
    ordinate is the $z$-coordinate of atoms along the $[001]$ direction
    of the crystal, which is normal to the interface.
    (a) $\Delta \theta \equiv$ RMS deviation of the nearest neighbor 
	bond angles from the ideal tetrahderal angle of $109.5^{\circ}$; 
    (b) $\left| \vec{v_t} \right| \equiv$ Magnitude of the sum of the 
	nearest neighbors vectors;
    (c) $\Omega_v \equiv$ Voronoi volume (volume of region closer to the 
	atom than to any other atom).  
    The letters `a' and `c' indicate the amorphous and crystalline
    regions of the samples, respectively.  The vertical dashed lines
	correspond to the position of the interface.  }
    \end{minipage}
    \end{center}

    \label{fig:local_order}
\end{figure}

The characterization of the interface is somewhat more demanding.  In
order to identify the interface region and to characterize its features
we define three different quantities.  The first of these is the RMS
deviation of the bond angles from the ideal tetrahedral angle
$\Delta\theta$.  The bond angle deviation for each atom vs. its
$z$-coordinate, averaged over all samples and smoothed by averaging
over a thickness $\Delta z = 1.0$~\AA, is plotted in
Fig.~\ref{fig:local_order}~(a).  Although the differences between the
crystal and amorphous regions are small (due to the strong angular
forces in silicon), a $7$~\AA\ thick transition region associated with
the interface is clearly visible between $z = 5$~\AA\ and $12$~\AA, and
between $z = 17$~\AA\ and $24$~\AA.  This observation is in contrast to
results of Spaepen~\cite{Spaepen78} from an analysis of a hand built
model for a $(111)$ interface relaxed with a Keating potential that
shows a larger bond angle deviation at the interface than at either of
the adjacent phases.

A second quantity we define to characterize the interface is the sum of
the vectors pointing from an atom to its nearest neighbors.  This
vector quantifies the asymmetry of the atomic environment.  For
example, if an atom is missing one of its neighbors while retaining
$sp^3$ bonding, this vector will point away from the missing atom. We
refer to this vector as the ``tetrahedral vector,'' $\vec{v_t}$.
Because of the difficulty of plotting vector quantities, the magnitude
of $\vec{v_t}$ vs. the $z$-position of each atom is plotted in
Fig.~\ref{fig:local_order} (b), averaged over all samples and smoothed
as described earlier.  The differences betweeen the crystalline and
amorphous regions are again small but distinct.  The extent of the
interface using $\vec{v_t}$ is very similar to that indicated by
$\Delta\theta$.  In the interface region, both $\Delta\theta$ and
$\vec{v_t}$ vary monotonically between the values in the amorphous and
the crystal regions.  The definition of the vector sum becomes more
useful when its values and directions at individual interface atoms are
considered:  these indicate the direction and amount by which a given
atom (or one of its neighbors) should move in order to create an
environment closer to the crystalline state.

A third local quantity we employed to characterize the interface region
is the volume of the Voronoi polyhedron associated with each atom
$\Omega_v$, plotted in Fig.~\ref{fig:local_order} (c), averaged over
all samples and smoothed as described earlier.  This quantity gives a
local measure of the density, as well as an estimate of the free volume
around each atom.  $\Omega_v$ is about 19.0~\AA$^3$ in the crystal,
which corresponds to a 3.5\% compression of the unstrained bulk crystal
volume.  In the amorphous region $\Omega_v$ ranges from 20.0~\AA$^3$ to
20.5~\AA$^3$, i.e. the amorphous phase is a few percent less dense than
the crystal.  To determine the variation of the strain with position
and direction, we calculated the mean nearest neighbor spacing
projected along the in-plane axes, $x$ and $y$, and the perpendicular
axis, $z$.  In the crystalline region the $x$ and $y$ spacings are 7\%
smaller than the spacing along the perpedicular direction, indicating
that the crystal is under biaxial compression.  In the amorphous region
the $x$ and $y$ spacings are 3\% larger than in the perpendicular
direction, indicating that the amorphous is under biaxial tension.
Because the two adjacent phases are in opposite strain states, it is
impossible to resolve the sign or magnitude of the interface stress.

\begin{figure}[p]
    \begin{center}
    \epsfig{file=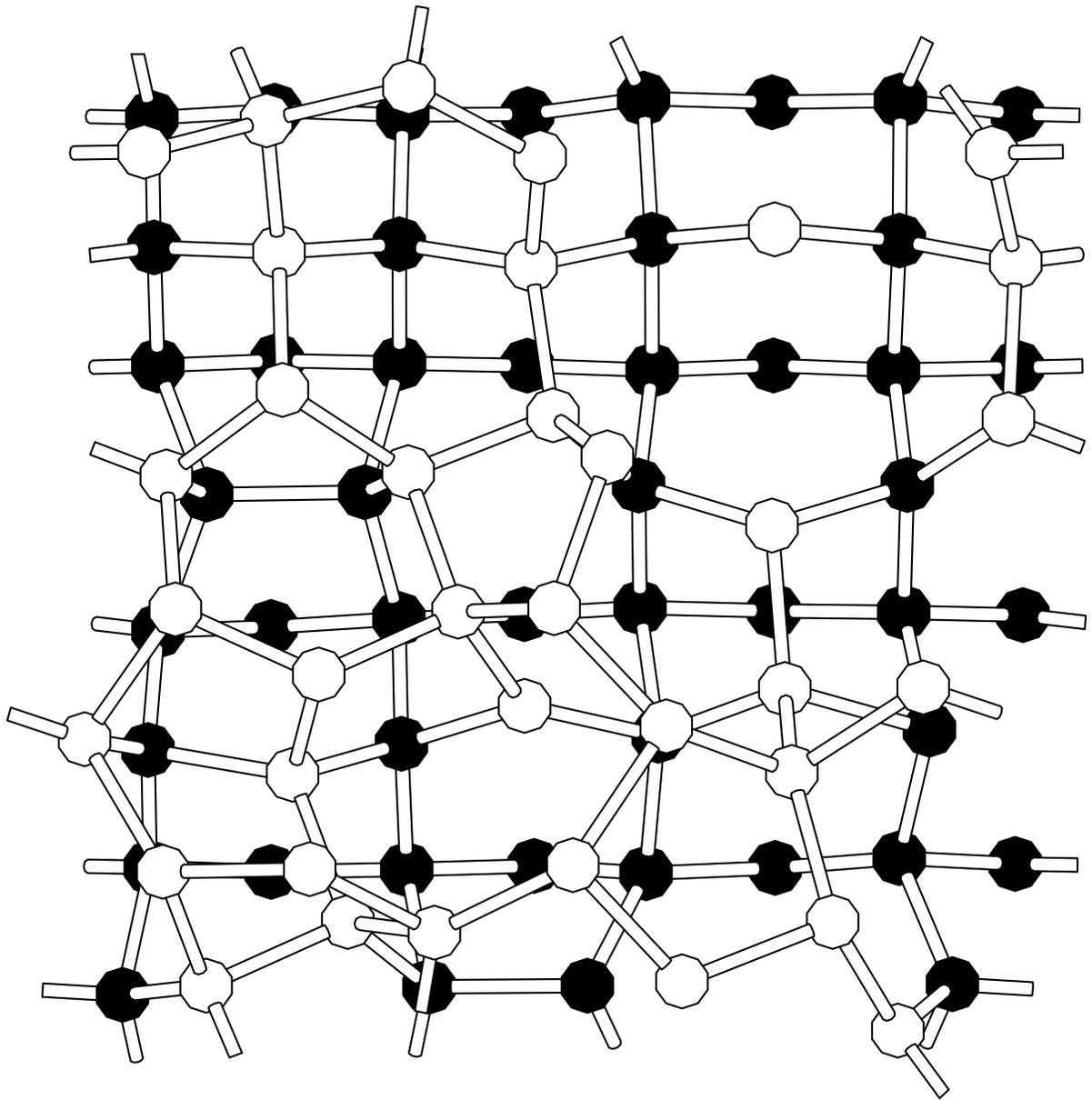,height=6.5in}
    \end{center}
    \begin{center}
    \begin{minipage}[t]{5in}
    \caption{Plan-view of an amorphous-crystal interface with the
    same conventions as in Fig. 1.  One dimer defect in the crystalline
    region near the bottom center of the image and one near the left
    center are easily seen. }
    \end{minipage}
    \end{center}

    \label{fig:pic_interface} 
\end{figure}

For a better understanding of the structure of the amorphous-crystal
interface we created slices of the samples parallel to the interface.
Perspective views of these slices reveal some interesting
characteristics:  Fig.~\ref{fig:pic_interface} is an example where the
prominent features of the crystalline portion are chains of atoms along
the $[110]$ direction, with very few defects.  The atoms that are not
in ideal positions form dimers, where pairs of atoms on adjacent
$[110]$ chains have come close together to form a bond, a feature which
was also seen in the hand-built model of Saito and
Ohdomari~\cite{SaitoOhdomari8184}.  One example of this defect is seen
on the left side of the image in Fig.~\ref{fig:pic_interface} (between
the two vertical $[110]$ chains).  This feature is very similar to the
well known Si(001) $2\times1$ free surface reconstruction, although in
the present case the atoms participating in the dimer have four bonds
(each with two more neighbors on the crystalline side and one more
neighbor on the amorphous side).  On the amorphous side of the
interface, some atoms are beginning to assume positions compatible with
the crystal lattice.  They line up in chains along $[110]$ directions
(center of image in Fig.~\ref{fig:pic_interface}), as would be expected
for the next layer in the crystal.  The remaining atoms are arranged in
more disordered configurations.

\section{Interface Energies}

One important quantity that characterizes the interface is the
interfacial tension $\sigma_{ac}$, which is, for a single-component
system, the excess free energy per unit area.  This excess is
responsible for the barrier to nucleation of the crystal in the middle
of the amorphous phase;  typically the interfacial tension is
determined experimentally by interpreting nucleation rate measurements
under conditions where heterogneous nucleation is believed to be
insignificant.  Because it is difficult to ensure that this condition
has been achieved, experimental values for the interfacial tension,
such as those estimated by Tsao and Peercy~\cite{Tsao87} or
Yang~\cite{YangThesis}, represent a lower limit on the true value of
the interfacial tension.  At sufficiently low temperatures the entropic
contribution to the interfacial tension can be neglected and
$\sigma_{ac}$ can be approximated by the excess interfacial energy per
unit area $\varepsilon_{ac}$, which is easier to determine
theoretically.  Mathematically, $\varepsilon_{ac}$ is defined as the
excess energy of a system with an interface over the weighted sum of
the energies of the two constituent phases,
\begin{equation}
    \varepsilon_{ac} = (E - N_c\varepsilon_c - N_a\varepsilon_a)/A
    \label{eq:epsilon_ac}
\end{equation}
$E$ is the total cohesive energy of the sample with the interface,
$\varepsilon_c$ and $\varepsilon_a$ are the cohesive energies per atom
of the reference crystal and amorphous states, $N_c$ and $N_a$ are the
number of atoms in the crystalline and amorphous phases, respectively,
and $A$ is the total area of the interface.  An analogous equation
to Eq.~(\ref{eq:epsilon_ac}) for $\sigma_{ac}$ can be obtained by replacing
$\varepsilon_{c}$ and $\varepsilon_{a}$ by the corresponding free
energies of these phases per atom, $g_{c}$ and $g_{a}$.  When the
system is in equilibrium, the assignment of atoms to the individual
phases (i.e., the determination of $N_c$ and $N_a$) is unnecessary
because, by definition, $g_{c}$ and $g_{a}$ are equal in equilibrium.
For the silicon amorphous-crystal interface, even when $\sigma_{ac}$
can be approximated by $\varepsilon_{ac}$, the determination of $N_{c}$
and $N_{a}$ is necessary because the two phases are not in equilibrium
with each other.  Hence we must determine which atoms should be
considered ``crystalline'' and which ``amorphous.''

To do that we visualize slices of our samples parallel to the interface
and label as crystalline any atoms that are bonded to two atoms that
were kept frozen or two other atoms that are labelled as crystalline by
this procedure, provided that the two atoms would share a common
neighbor in the perfect crystal.  This ensures that all the atoms that
are considered part of the crystal are in a nearly ideal crystal
environment on at least one side, and all are members of six-fold rings
that are contained in the crystal.  Because the calculated interface
energy is sensitive to the number of crystal atoms we need to employ a
more rigorous definition of the bond between atoms than the one used
earlier, which relied simply on distance (atoms closer than
2.7~\AA\ were considered bonded).  To this end, we consider atoms
bonded only if the tight-binding charge density half way between them
is above a threshold value which is obtained by using representative
$s$ and $p$ orbitals attached to each atom.  Typically, between 10 and
20 pairs of neigboring atoms (out of about 650 pairs in each sample)
have charge densities that fall below this threshold and are not
considered to be bonded to each other, even though their distance
is shorter than 2.7~\AA.

A second complication in using our tight-binding Hamiltonian to compute
the interface energy is the precise value of $\varepsilon_c$ and
$\varepsilon_a$.  The reference crystal state is an uncompressed
diamond lattice (the compression energy is negligible), trivial to
generate and its cohesive energy $\varepsilon_c$ is easy to compute.
To compute an appropriate reference amorphous state from which
$\varepsilon_a$ can be estimated, we take each interface sample and
apply the same procedure we used to create the amorphous portion, but
this time keeping a 4.75~\AA\ slab centered in the middle of the
amorphous portion frozen.  In this way, we turn the entire sample
amorphous.  Each bulk amorphous sample is then relaxed with the
tight-binding Hamiltonian, and used as the reference state when
computing the interface energy for the corresponding interface sample.
The resulting bulk amorphous samples have cohesive energies
$\varepsilon_a$ between 4.519 and 4.536~eV/atom, corresponding to an
excess energy for the amorphous phase $\Delta \varepsilon_{ac}$ of 0.17
to 0.19~eV/atom.  These values are a factor of two higher than an
experimental value for $\Delta \varepsilon_{ac}$ of 0.097~eV/atom, as
extrapolated to 0~K from Donovan's measurement~\cite{Donovan85} at 960K
using the specific heat listed in that work.

The resulting interface energies $\sigma_{ac}$ range from 0.39 to
0.54~J/m$^2$ for the six different samples, with a mean of 0.49~J/m$^2$
and a standard deviation of 0.05~J/m$^2$.  The scatter is due to several
factors.  The total energy of the two interfaces in each sample is a
small number (about 15~eV) computed by subtracting large numbers (total
energies for the interface and reference states, each of order
1500~eV).  Scatter of 0.3\% in the total energy of the interface
samples or reference amorphous samples (which is inevitable due to
their disordered nature and small size of the systems) causes a scatter
of 30\% in the computed interface energy.  Partitioning the atoms into
crystalline and amorphous parts also involves an error of about two or
three atoms per interface, arising from both the threshold charge
density value for considering two neighboring atoms bonded and from
errors made in the manual counting process.  There is also a
potentially larger source of error in the arbitrary definition of what
is required for an atom to be considered ``crystalline.''  Some other
criteria we considered, using the values of different measures of order
to distinguish between ``crystalline'' and ``amorphous'' atoms, gave
values for $N_c$ that differed by as many as tens of atoms from the
topological criterion described previously.

The only previous attempt to compute the interface energy through
simulation we are aware of is Spaepen's work~\cite{Spaepen78} using a
Keating potential to evaluate the energy of each atom in a hand built
model of a $(111)$ interface;  the computed interface energy was
0.31~J/m$^2$.  Saito and Ohdomari~\cite{SaitoOhdomari8184} also
computed the Keating potential energy as a function of distance from
the interface, although they did not publish a corresponding interface
energy.  Using their plot of the excess energy, and considering their
``original surface'' as a part of the crystal, we compute an interface
energy of 0.23~J/m$^2$.  These values are consistent with our
calculation considering the substantial differences in interface
geometry and computational methods.  The most recent experimental
measurement of the amorphous-crystal interface tension for silicon we
are aware of is by Yang~\cite{YangThesis}: an interface tension of
0.48~J/m$^2$ was obtained by fitting a physically motivated kinetic
model to the observed nucleation rate of crystals during ion-beam
enhanced crystallization of an amorphous sample.  The agreement of this
value with our calcalution is excellent, but probably fortuitous.  The
only other experimental result we are aware of is the work by Tsao and
Peercy~\cite{Tsao87}.  They deduced an interface tension of
0.04~J/m$^2$ from K\"oster's nucleation rate measurements for amorphous
thin films~\cite{Koster78}, where the nucleation is unlikely to be
homogenous, and is therefore not a reflection of the true interface
tension.

\section{Summary}

Using a combination of interatomic potentials and a specially optimized
nonorthogonal tight-binding Hamiltonian we have created
amorphous-crystal interfaces in silicon by performing melt and quench
numerical experiments.  The interfaces are about 7~\AA\ thick.  All
measures of order we considered interpolated smoothly between the
crystalline and amorphous values.  Slices of the sample along the
interface reveal features analogous to dimers on the Si(001) surface
and short crystal-like chains in the amorphous layer adjacent to the
crystal. By comparing the energies of samples with and without
interfaces we compute an interface energy of about 0.49~J/m$^2$, in
good agreement with experimental evidence and other theoretical work.

\acknowledgements

This research was supported by the Harvard MRSEC under NSF-DMR-94-00396.

\end{multicols}

\begin{table}
    \begin{center}
    \begin{minipage}[t]{5in}

    \caption{Coordination and ring statistics averaged over six
    samples.  Coordination statistics are tabulated separately for the
    crystalline and amorphous regions.  Note that the rings are too
    large compared to the thickness of the crystalline region to allow
    for such a separation, so values averaged over the entire sample
    are listed.  }

    \begin{tabular}{lrrrrrr}
	\multicolumn{5}{c}{Coordination Statistics} \\
        Neighbor Num.           & 2     & 3     & 4      & 5   \\
        crystal         	& 0.1\% & 0.4\% & 98.6\% & 0.9\% \\
        amorphous       	& 0.1\% & 2.1\% & 94.5\% & 3.3\% \\
	\multicolumn{5}{c}{Rings per Atom} \\
	Ring Size 	& 3 &     4 &     5 &     6 &     7 &	8\\
	all rings	& 0.01 & 0.04 &  0.36 & 1.11 &  0.86 & 2.59  \\
	minimal rings	& 0.01 & 0.04 &  0.36 & 0.99 &  0.25 & 0.01 
    \end{tabular}
    \end{minipage}
    \end{center}

    \label{table:bulk_stats}
\end{table}

\end{document}